\documentclass[prb,twocolumn,showpacs,preprintnumbers,amsmath,amssymb,aps]{revtex4}

\usepackage{graphicx}
\usepackage{dcolumn}
\usepackage{epsf}
\usepackage{bm}

\begin{document}


\title{ Large Orbital Magnetic Moment and  Coulomb Correlation effects in FeBr$_2$} 

\author{ S. J. Youn, B. R. Sahu and Kwang S. Kim}
\affiliation{%
     National Creative Research Initiative Center for Superfunctional 
Materials and Department of Chemistry, Pohang University of Science and 
Technology, Pohang 790-784, South Korea.
}
\date{\today}

\begin{abstract}
   We have performed an all-electron fully relativistic density functional
calculation to study the magnetic properties of FeBr$_2$. 
We show for the first time that   
the correlation effect enhances the contribution 
from orbital degrees of freedom of $d$ electrons 
to the total magnetic moment on Fe$^{2+}$ as opposed to 
common notion of nearly total quenching of the orbital moment on Fe$^{2+}$ site.
The insulating nature of the system is correctly predicted when the Hubbard 
parameter U is included. 
Energy bands around the gap are very narrow in width and originate
from the localized Fe-3$d$ orbitals, which indicates that
FeBr$_2$ is a typical example of the Mott insulator.
\end{abstract}

\pacs{ 74.25.Ha, 71.27.+a, 71.20.-b}
\maketitle

   FeX$_2$(X= Cl, Br, I) are well known {\it  model} systems for the 
study of antiferromagnetism.\cite{stryjewski}  They show unusual 
{\it metamagnetic} behavior wherein they undergo a phase 
transition from the antiferromagnetic to a saturated paramagnetic phase
as a function of external magnetic field and temperature.
FeBr$_2$ has a layered structure, where the ferromagnetic Fe layers
are widely separated by non-magnetic halogen layers producing 
a quite weak interlayer antiferromagnetic interactions ($T_N$= 14.2 K).
The long range antiferromagnetic ordering can be overcome by applying an 
external  magnetic field ($\sim$ 30 kOe) parallel to the c-axis.\cite{yelon}
As a result, many theoretical \cite{pleming} and
experimental\cite{binek,aruga} works on FeBr$_2$ have been
concentrated on probing the
axial magnetic phase diagram, to predict and understand anomalies near
the antiferromagnetic to the paramagnetic phase boundary.

In addition, the electronic structures of transition metal dihalide system
are of interest because they show various types of interesting 
behavior due to the strong correlation effect in the transition metal 
ion.\cite{zaanen}
For example, they behave as a Mott insulator or charge transfer type insulator 
depending upon the relative size of charge transfer energy $\Delta$ and
the intra-orbital correlation energy $U$ of the $d$-electrons.

Although large amount of work is done to understand the anomalies in
the phase diagram and the correlation effects,
surprisingly no electronic structure 
is known from the first principles calculation.
In this report, we study the electronic and magnetic properties of 
one of these systems (FeBr$_2$) by
combining a first principles method and the Hubbard on-site correlation term.

  Orbital magnetic moment plays a crucial role in determining many important
effects in magnetic materials such as magnetic crystalline anisotropy, 
non-collinear magnetism, magneto-optical Kerr effect {\it etc.}.\cite{ebert}
It is a well known fact that crystal field interaction quenches the 
orbital magnetic moment in systems containing 3d-transition metal.
Accordingly, it is assumed that because of strong crystalline field on 
Fe$^{2+}$ ions from the surrounding Br$^-$, the orbital moment 
will also get quenched in FeBr$_2$.\cite{wilkinson}
Ropka, Michalski, and Radwanski\cite{ropka}  showed that the orbital moment is
not quenched fully by the crystal field through their quasi-atomic calculation.
Ropka {\it et al.} ascribed the origin of the large orbital moment to 
spin-orbit coupling (SOC).
Our study also demonstrates that orbital degrees of freedom has substantial
contribution (of $\sim 20 \%$ to the total magnetic moment). 
However, correlation effect in Fe 3$d$ orbitals increases the orbital 
contribution to the total magnetic moment significantly,
as discussed below.  This is explicitly shown by
incorporating the Hubbard parameter U in the self-consistent calculation
which also predict an insulating nature.

\begin{figure}
\epsfysize 8cm
\epsfbox{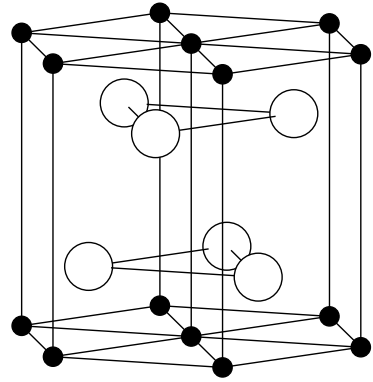}
\caption{Crystal structure of FeBr$_2$. Black and white spheres
represent Fe and Br atoms, respectively.}
\label{fig1}
\end{figure}

         FeBr$_2$ is isomorphic to the hexagonal CdI$_2$ structure
(space group  $p{\bar {3}}m1$, $D^3_{3d}$, No. 164) at 
atmospheric pressure and low temperature. 
Figure \ref{fig1} shows the crystal structure.
The lattice parameters are 
{$a = 3.77$ \AA}  and $ c = 6.22$ \AA\cite{wilkinson}
where the $c/a$ ratio ($\sim 1.65$) is very close to 
the ideal value of $\sqrt{8/3}\sim 1.63$. 
The chemical unit cell has one Fe and two Br atoms. 
The Br atoms are bonded tightly to the Fe 
atoms on either side by covalent bond to form a Br-Fe-Br trio.
The Br-Fe-Br trios are 
separated by an empty layer between them,
and bonding between the trios is a weak van der Waals type.
The Fe ions are located at 1{\it a} positions whereas
Br atoms are located at 2{\it d} positions in a unit cell. Each Fe ion 
is at the center of octahedron formed by the surrounding Br ions
with a small distortion. As a 
result, a small trigonal component of crystal field due to this distortion is 
present apart from the cubic component.  
 FeBr$_2$ is also taken as an example of dynamic Jahn-Teller system
wherein strong coupling between lattice and spin excitations are manifested
in temperature dependent Raman spectrum.\cite{johnstone}

\begin{table}[b]
\caption{ Direct energy gap(E$_g^d$), Indirect energy gap(E$_g^i$), orbital magnetic moment($\mu_{l}$), spin magnetic
moment($\mu_{s}$), and total magnetic moment $\mu$ = $\mu_{l}+ \mu_{s}$ for 
FeBr$_2$. The energy gap is in eV and magnetic moments are in Bohr magneton. 
U= 0 eV in LDA calculation.
\label{tab1}
}
\begin{ruledtabular}
\begin{tabular}{ccccc}
          & Exp. & LDA(U=0) & U= 5.7& U= 6.0  \\
 \hline
 E$_g^d$     & 2.0\footnotemark[1] & 0.0 & 1.76 & 1.87     \\
 E$_g^i$     &         & 0.0 & 1.43 & 1.53     \\
 $\mu_{l}$ & & 0.14 & 0.66  & 0.66  \\
 $\mu_{s}$ & & 3.50& 3.85 &  3.86  \\
 $\mu$     & 4.4\footnotemark[1] & 3.64 & 4.51 &  4.52  \\
\end{tabular}
\end{ruledtabular}
\footnotetext[1]{Reference [9]}
\footnotetext[2]{Reference [8]}
\end{table}

          We have carried out 
density functional calculations\cite{hohenberg} on FeBr$_2$ using the 
full-potential linear muffin-tin orbital (FPLMTO)\cite{savrasov} method  
with full relativistic effects on the single-particle electron states
and the nonrelativistic local density approximation (LDA) for 
the many-body effects in the exchange-correlation energy with the 
Vosko-Wilk-Nussair form.\cite{vosko}
The strong correlation effect in the Fe 3$d$ site is taken into account
by LDA+U method.\cite{ldau}
In the past, LDA+U method is frequently used to predict the occurrence of large 
orbital moment and insulating nature and is found to be a robust
method for the correlated systems.\cite{solovyev,min}
The chemical unit cell is doubled ($c = 12.44$ \AA) along 
c-axis to simulate
the antiferromagnetic system with 2 Fe atoms and 4 Br atoms. 
Basis functions, electron 
densities and potential were calculated without any shape approximation.
The trial wave function, potential and charge densities are  
expanded in combination of spherical harmonics (l$_{max}$ = 6) inside the
atomic muffin-tin sphere region and plane wave
Fourier series in the interstitial region (E$_{cut}$=94 Ry). 
2 $\it \kappa$-basis set (where $\it \kappa$ is kinetic energy of muffin-tin
orbitals in interstitial region) is used.  
In the Fe sphere, 4{\it s}, 4{\it p} and 3{\it d} orbitals are taken as valence 
states and 3{\it p} orbitals are treated as semi-core
state, whereas in the Br sphere, 
{\it 3d} orbitals are taken as a semicore state and 
4{\it s}, 4{\it p} and 4{\it d} orbitals are taken as valence states. 
The semicore states are treated in
a separate energy window. 
The {\bf k}-space integration is done with
30 {\bf k}-points
in irreducible part of Brillouin Zone(IBZ)(144 {\bf k}-points in full Brillouin
Zone(FBZ))  
using the tetrahedron method.\cite{blochl}
We have also carried out calculations with larger set of
{\bf k}-points (43 points in IBZ and 320 points in FBZ), 
but the total energies 
are found to be converged within 1 meV and magnetic moments change within 0.01
$\mu_{B}$. To bring about the orbital contribution to total magnetic
moment, spin-orbit interaction 
is included in the crystal Hamiltonian.
The calculations are done at experimental lattice constant. 

   Due to mean field character of LDA, it incorrectly 
describes the ground state of many strongly correlated materials, for example,
the transition metal oxide systems.\cite{anisimov} In these
materials, the $d$-orbitals are well localized and retain the strong
atomic-like character. For a good description of strong on-site correlation
effect between electrons in the $d$-shell, the LDA+U method
is widely used.  This method identifies these orbitals as {\it correlated}
states.  The LDA+U variational total energy functional takes the form

\begin{eqnarray}
E^{tot}[\rho, \hat{n}]=E^{LDA}[\rho]+E^{U}[\hat{n}] - E^{dc}[\hat{n}],
\end{eqnarray}

     where the first term is the usual local spin density functional of 
the local electron spin density $\rho^{\sigma}$({\bf r})($\sigma = \uparrow, \downarrow $). $\hat{n}^{\sigma}$ is a local orbital ($d$ or $f$) occupation 
matrix. E$^{U}$ is an electron-electron interaction energy which depends
on Slater integrals F$^{0}$, F$^{2}$ and  F$^{4}$.
E$^{dc}$ is a double
counting term. The Slater integrals are in turn related to the {\it screened}
Coulomb and exchange parameters U and J respectively:

\begin{eqnarray}
J = \frac{F^{4}+F^{2}}{14}
\end{eqnarray}

and

\begin{eqnarray}
U = F^{0}.
\end{eqnarray}

For 3$d$ electron systems it is found that $F^2/F^4$ = 0.625.
It is to be noted that the orbital polarization is automatically 
included in LDA+U method, as the orbital dependent effect comes from F$^{2}$ 
and F$^{4}$. 
Since no experimental value of U in 
FeBr$_2$ system is available in the literature,
we decided to take U and J values from experiments on other systems.
It is to be noted that the experimental value of U in 
FeO system\cite{tanaka} is 5.7eV and 6.0 eV.
In similar systems like 
antiferromagnetic transition metal oxides,\cite{ldau}
J is close to 0.9 eV and does not change much from this value,
so we have taken J = 0.9 eV.
To check the validity of our U values,
we performed the self-consistent calculations 
at four different values of U, namely, 5.7 eV, 6.0 eV, 6.4 eV and 6.8 eV. 
We consistently obtain the insulating ground state in agreement with the 
experiment, although the energy gap changes by a small fraction.
The magnetic moments are almost  constant with respect to the U values.
So, the four different values of U do not affect our conclusions.
We present our results for U = 6 eV taken from the experimental value 
in FeO system if not mentioned explicitly.

Table \ref{tab1} shows energy gap and magnetic moment for
FeBr$_2$ at the two representative values of U.
LDA and GGA both predict FeBr$_2$ to be metallic,
while LDA+U predicts it to be insulating in agreement with 
experiment.\cite{wilkinson}
It is, therefore, crucial to take into account 
the correlation effects to obtain the correct electronic structure. 
It means that there is a strong intra-orbital correlation for Fe$^{2+}$ ion 
in FeBr$_2$ and that this strong correlation is responsible for the 
insulating nature.

\begin{figure}
\epsfysize 8cm
\epsfbox{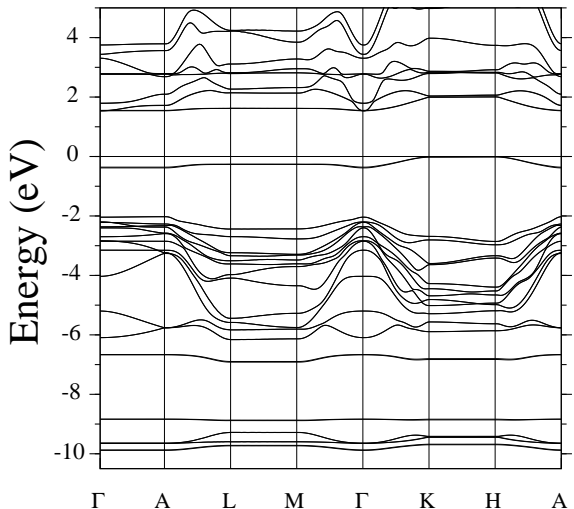}
\caption{ 
Energy band structure of FeBr$_2$ at U = 6.0 eV along high symmetry directions 
in IBZ of the hexagonal unit cell. 
The top of the valence band is set to energy zero.
}
\label{fig2}
\end{figure}

Figure ~\ref{fig2} shows dispersion curves for electronic states along high 
symmetry directions in IBZ of the hexagonal unit cell.
The top of valence band is set to energy zero.
Br-4$s$ bands are not shown in the figure which is located at around -16 eV.
It is seen from Fig.~\ref{fig2} that
there is a indirect energy gap between H and $\Gamma$ with a 
gap size of 1.53 eV.
A direct gap  of 1.87 eV is located at the middle of 
the symmetry line connecting M and $\Gamma$.
No experimental value of the gap is available
in the literature. 
However, since FeBr$_2$ is yellow in color,\cite{ropka} the 
gap would be around 2 eV which may correspond to a transition across 
the direct gap.
Our calculated gap values of 1.87 eV are in reasonable agreement
with the conjectured one(see Table \ref{tab1}). 
   We can see from Fig. \ref{fig2} that the bands have little dispersion 
in the $z$-direction (see the bands along symmetry lines 
$\Gamma$-A, L-M, and K-H)  
which is a characteristic feature 
in systems with layered structure like graphite or some 
high T$_c$ superconductors.
The equienergy surfaces near the energy gap, therefore,  
have cylinder-like shapes.

\begin{figure}[h]
\epsfysize 10cm
\epsfbox{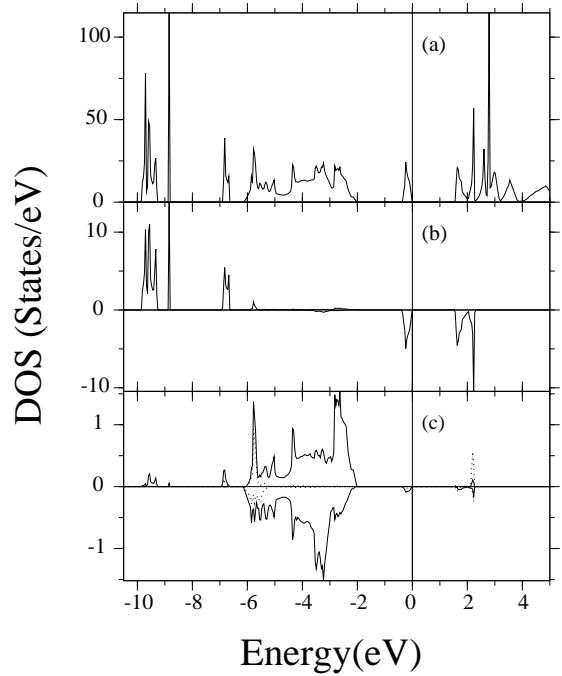}
\caption{ 
Density of states for FeBr$_2$. (a) Total DOS (b) Partial DOS of Fe-3$d$ states
(c) Partial DOS of Br-4$p$ ( solid line) and Fe-4$s$ ( dotted line). 
In (b) and (c), positive and negative DOS represent spin-up and spin-down DOS,
respectively.
}
\label{fig3}
\end{figure}

Density of states (DOS) of FeBr$_2$ 
in Fig. \ref{fig3}(a) shows the total DOS for FeBr$_2$.
Spin-up Fe-$d$ states are located deep in energy between
-10 eV and -6.5 eV, whereas
spin-down Fe-$d$ states are located at between -0.5 eV and 3 eV
as shown in Fig. \ref{fig3}(b).
Fe-$d$ orbitals form localized states 
with narrow band width where
the main peaks are located 
-9.6 eV, -8.9 eV, -6.8 eV for spin-up states 
and -0.2 eV, 1.6 eV, 2.2 eV for spin-down states.
A group of bonding orbitals with wide band width 
is located between -6 eV and -2 eV which
is the hybridization of Br-$p$ and Fe-$s$ bands( Fig. \ref{fig3}(c)).
This wide band gives a covalent bonding within the Br-Fe-Br trio.
The energy gap between 
the hybridized bonding orbitals of Br-$p$/Fe-$s$ 
and 
the conduction band is 3.6 eV.
The $d$-bandwidth, W, of 0.5 eV is found to be much smaller than U so that 
$ U/W \/> 1$. 
The states near the band gap 
at conduction band minimum and valance band maximum are mainly from spin-down
Fe-$d$ states. 
This means that FeBr$_2$ is a typical Mott insulator  
(materials in which the kinetic energy gain is smaller than the Coulomb 
repulsion energy U and as a result 
electrons can hardly hop to the Fe-$d$ orbitals). 

     Fe$^{2+}$ ion in FeBr$_2$ has six electrons in $d$ orbitals. 
    The nearly octahedral arrangement of Br$^-$ surrounding Fe$^{2+}$ ions
introduces crystal field which splits the d-orbital into t$_{2g}$ 
(d$_{xy}$, d$_{yz}$, and d$_{xy}$) and  e$_g$ (d$_{x^2 -y^2}$ 
and d$_{3z^2 -r^2}$) orbitals. 
This splitting hinders the free rotation of the 
electrons and reduces its orbital moment. 
Both t$_{2g}$ and e$_g$ orbitals of spin-up Fe-$d$ bands are fully occupied 
while only one of t$_{2g}$ orbitals is occupied in spin-down bands
to contribute 4 $\mu_B$ as a spin contribution to the total magnetic moment.
This simple estimation of the magnetic moment from DOS 
is consistent with the calculated 
spin contribution of 3.86 $\mu_B$ to the total magnetic moment.
The Fe$^{2+}$ ion has L=2, S=2 and J=4 in free space. 
If we assume that the orbital moment is quenched as in usual 3$d$ transition
metal magnetic system, the spin contribution to the total magnetic moment is 
also consistent with the simple atomic picture.
However, the neutron
diffraction study\cite{wilkinson} shows that the magnetic moment on Fe site is 
nearly 4.4 $\mu_{B}$. Wilkinson {\it et al.}\cite{wilkinson}
argue that this moment is close to that 
expected for divalent metallic Fe$^{2+}$ ions if the orbital contribution 
is taken to be zero.
   Our calculation shows that total magnetic moment $\mu$ is
split into orbital $\mu_l$ and spin parts $\mu_s$ with the orbital part 
contributing a significant amount. 
From Table \ref{tab1} it is seen that the orbital 
contribution to the 
total magnetic moment is $\sim 20 \%$ the total magnetic moment. It is 
interesting to note that the nearly octahedral arrangement of Br ions 
surrounding Fe ion is not able to totally quench the orbital degrees of 
freedom. 
The orbital and spin degrees of freedom are therefore not 
totally decoupled (obeying Hund's third rule). Quasi-atomic 
calculations by Ropka et al\cite{ropka} also give the orbital moment close to 
our values, and the large orbital moment 
were speculated to be due to SOC.  However, our calculations show that 
SOC is not enough to explain the large 
orbital contribution as can be seen in table \ref{tab1}.
LDA calculation with SOC
gives an orbital magnetic moment of 0.14 $\mu_B$ and total magnetic
moment of 3.64 $\mu_B$ which is smaller than the experimental
value of 4.4 $\mu_B$.
When we include the on-site correlation effect at the Fe site as well as 
SOC, we get 
about five times larger orbital magnetic moment of 0.66 $\mu_B$.
Total magnetic moment of 4.52 $\mu_B$ is in good agreement with experimental
value within 3\% error.
Thus we can see that the important contribution to the large orbital
magnetic moment of FeBr$_2$ is mostly due to
the correlation effect of the 3$d$ orbitals.
Therefore, the correlation effect 
plays a very important role in  determining the magnetic properties 
as well as the insulating ground state.
In Ropka {\it et al.}'s quasi-atomic model,
the correlation effects are included by the correlated basis set
with no explicit term representing on-site correlation effect
in the Hamiltonian.
However, in our calculation we include explicitly the correlation
term in the Hamiltonian.

    We suggest that, if an x-ray magnetic circular dichroism (XMCD)\cite{thole} 
experiment, which is a local probe 
for spin and orbital magnetic contribution to total magnetic moment, is 
performed on FeBr$_2$, then the orbital contribution can easily 
be probed. Such experiments relate, 
via atomic sum rules, the 
dichroic intensities ({\it ie.,} difference between right- and left-circularly
polarized photons absorption cross-sections) measured at the absorption edges of the constituent elements
to the  ground-state expectation value of effective one-electron operators such
as $\langle L_z \rangle$ and $\langle S_z \rangle$.

    To conclude, we have shown for the first time that 
the correlation effect is important 
in describing the magnetic and electronic properties of FeBr$_2$ 
which plays an important role in enhancing the orbital magnetic moment
and predicting the correct insulating nature.
This indicates that Fe$^{2+}$ ion in FeBr$_2$ retains to some
extent its atomic-like character. 
FeBr$_2$ can be classified as a typical example of Mott insulator
due to the character of the energy gap and 
the width of the bands around the gap.
We expect that the present 
calculation will help understand the electronic and magnetic properties of other
iron halide systems. 

     We are thankful to Prof. B. I. Min and Prof. Sukmin Jeong for 
helpful discussion. Critical comments from Prof. D. I. Khomskii is
highly appreciated. The authors are grateful to KISTEP/CRI of 
Korean ministry of Science and Technology for financial support.

\end{document}